# LAGOS-AND: A Large Gold Standard Dataset for Scholarly Author Name Disambiguation

Li Zhang, School of Information Management, Wuhan University, 299, Bayi Street, Wuchang District, Wuhan, China. Email: zlahu@foxmail.com

Wei Lu, School of Information Management, Wuhan University, 299, Bayi Street, Wuchang District, Wuhan, China. 027-68754270. Email: weilu@whu.edu.cn

Jinqing Yang, School of Information Management, Wuhan University, 299, Bayi Street, Wuchang District, Wuhan, China. Email: Jinq_yang@163.com

Correspondence: Prof. Wei Lu, Ph.D. School of Information Management, Wuhan University.

**Abstract**

In this paper, we present a method to automatically build large labeled datasets for the author ambiguity problem in the academic world by leveraging the authoritative academic resources, ORCID and DOI. Using the method, we built LAGOS-AND, two **la**rge, **go**ld-**s**tandard datasets for author name disambiguation (**AND**), of which LAGOS-AND-BLOCK is created for clustering-based AND research and LAGOS-AND-PAIRWISE is created for classification-based AND research. Our LAGOS-AND datasets are substantially different from the existing ones. The initial versions of the datasets (v1.0, released in February 2021) include 7.5M citations authored by 798K unique authors (LAGOS-AND-BLOCK) and close to 1M instances (LAGOS-AND-PAIRWISE). And both datasets show close similarities to the whole Microsoft Academic Graph (MAG) across validations of six facets. In building the datasets, we reveal the variation degrees of last names in three literature databases, PubMed, MAG, and Semantic Scholar, by comparing author names hosted to the authors' official last names shown on the ORCID pages. Furthermore, we evaluate several baseline disambiguation methods as well as the MAG's author IDs system on our datasets, and the evaluation helps identify several interesting findings. We hope the datasets and findings will bring new insights for future studies. The code and datasets are publicly available.

# INTRODUCTION

Author name ambiguity is a well-known issue in academic literature databases/digital libraries. The name ambiguities in the real world reflect authors represented by name variants (synonyms), and some authors share the same name (homonyms) (Aman, 2018; Kim & Kim, 2020; Shoaib et al., 2020). This problem is very challenging in literature databases because, for example, there were about 40K citations[1] authored by "Wei Wang" in Microsoft Academic Graph (MAG) as of March 2019, and the name ambiguity problem is even more pronounced for the abbreviated names. Although some databases such as MAG and AMiner[2] have provided disambiguated author identifiers (IDs), the performance of the created author ID systems based on author name disambiguation (AND) approaches for million-scale databases is far from satisfactory (Zhang et al., 2020).

Identifying author uniqueness is crucial for many studies and applications. For instance, in the field of bibliometric research, a recent high-impact study used disambiguated author IDs to meet a larger goal of examining gender inequality in scientific careers (Huang et al., 2020). In digital library management research, Zhang et al., (2018) claimed that AND is a core component of AMiner, which is a free online service for academics.

To address the name ambiguity problem, the research community has developed many labeled AND datasets in recent years (see supplemental material A for the list of the AND datasets) to help develop supervised or semi-supervised disambiguation methods (Louppe et al., 2016; Mihaljevic & Santamaría, 2021), as well as test the performance of various disambiguation methods (Kim & Kim, 2020; Tekles & Bornmann, 2020). However, we find that existing datasets suffer from several issues or limitations. To be specific, the issues or limitations are as follows. (1). Unclear dataset creation process. Most datasets such as GS-PubMed (Vishnyakova et al., 2019) and SCAD-zbMATH (Müller et al., 2017) are created manually, meaning that many annotators are involved and a great deal of effort has to be invested. In addition, for some datasets such as Han-DBLP (Han et al., 2005), the details of creation are not thoroughly described, e.g., the quality assurance measures, which may raise concerns about the quality of the datasets. (2). Limited scale. Existing datasets are mostly limited in size (see the supplemental material A); however, in literature databases, the name ambiguity problem is generally more complex than that in small datasets (Xiao et al., 2020). This problem may make the data-driven disambiguation methods perform poorly in real literature databases. (3). Limited number of unbiased datasets. Existing datasets are unable to reach the level of the gold standards. One example that indicates the biases is that most datasets are designed to address the name homonym problem (Sanyal et al., 2021); however, the name ambiguities include synonyms as well as homonyms.

All of these issues and limitations not only bias the disambiguation results but, more importantly, hinder the development of effective AND methods. Motivated by the credibility and increasing popularity of the two academic resources Open Researcher Contributor Identification[3] (ORCID) and Digital Object Identifier[4] (DOI), we herein propose a method to automatically build improved datasets for AND. Our proposed method and the created datasets can address the above issues or limitations appropriately. Specifically, (1). Because ORCID iDs and DOIs are able to identify authors and scientific papers unambiguously, the publication history of an author (query DOIs by ORCID iD) and the authorship of a paper (query ORCID iDs by DOI) can be accurately identified. Thus, relying on this information it is feasible to develop an automatic method that can effortlessly build AND datasets. In addition, as the rationale of the developed method is simple and clear, it is also feasible to regenerate the datasets and create a new version of the AND datasets. (2). The two academic resources have gained increasing popularity in recent years. As shown in Figure 1, the two databases have been growing at a high and constant speed recently, and the number of ORCID records and DOI records in 2018 was 1,561,789 and 5,020,071 respectively. Therefore, such a large amount of labeled data made it possible to build a large AND dataset. (3). The proposed method is manageable and controllable due to its simplicity, meaning that, by adjusting the created datasets and comparing them with a real literature database, we can improve the quality of the datasets in several aspects, such as the covered ambiguity patterns.

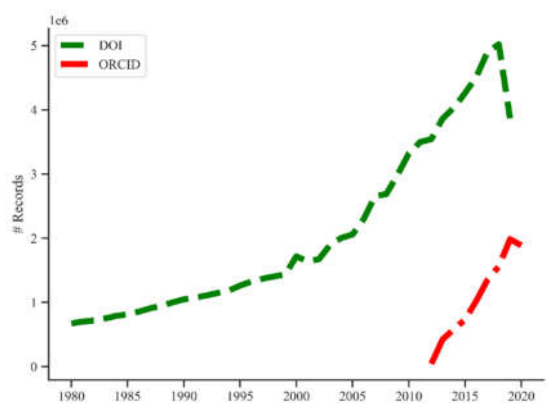

Figure 1. The number of ORCID and DOI records each year

In summary, our contributions are as follows:

- We develop a method that can automatically build large labeled datasets for the author name disambiguation research. The method is clearly presented and can be reused to generate new versions of the datasets.

- We have built LAGOS-AND based on the proposed method, which are two **la**rge, **go**ld-**s**tandard datasets for **AND**. To the best of our knowledge, our datasets are the world's largest AND datasets. The technical validation demonstrates that the two datasets show close similarities to the whole MAG across validations of six facets. The initial versions of our datasets are available at https://zenodo.org/record/4568624.

- We calculate the degree of last name variation in building the datasets. Evaluation results for three large literature databases show that the degrees range from 5.80% to 6.34%, and the variation degrees are even higher if (1) a popular name-parsing tool is used to extract the last names from full names for name comparison or (2) the accented alphabets are not transliterated to the standard characters (e.g., "á" → "a").

- We evaluate several baseline disambiguation methods and the author ID system of MAG on our datasets. The experimental results indicate that MAG's author IDs show poor performance on the two gold standards and that incorporating a semantic relatedness feature of citations boosts the performance of disambiguation. The code is available at https://github.com/carmanzhang/LAGOS-AND.

## RELATED WORKS

In this section, we review the most important datasets created for AND research[5]. According to our survey, there are at least 12 datasets available so far. These datasets have been widely adopted to develop disambiguation methods in various scenarios or for different objectives. However, we have identified a number of unresolved and even undiscovered issues with the datasets.

First, most of the existing datasets were created manually (Han et al., 2005; Vishnyakova et al., 2019; Xiao et al., 2020), which means that a great deal of effort needed to be invested in the data creation process. For example, a large group consisting of 22 annotators was involved in creating a dataset for AMiner (Wang et al., 2011), and each publication was annotated by at least three annotators to ensure a high annotation accuracy. Some datasets were created in a crowdsourcing fashion on platforms such as Amazon Mechanical Turk (MTurk). The method of building labeled datasets is usually considered effective (Zhang et al., 2016); however, the method appeared to be ineffective when it was used to create AND datasets, as a recent study (Vishnyakova et al., 2019) found that it was hard to control the data quality because the annotation tasks were distributed to many untrusted annotators who may try to guess the class labels rather than find the ground truths.

Second, current datasets are either limited in size or limited in scope. As shown in supplemental material A, most datasets contain fewer than 10,000 citations. However, small datasets such as Han-DBLP (Han, Xu, et al., 2005; Han, Zha, et al., 2005), Qian-DBLP (Qian et al., 2015), Kim-DBLP (Kim, 2018), Tang-AMiner (Tang et al., 2012; Wang et al., 2011), Culotta-REXA (Culotta et al., 2007), Cota-BDBComp (Cota et al., 2010), Song-PubMed (Song et al., 2015), and GS-PubMed (Vishnyakova et al., 2019) may be unable to adequately reflect the real complexity of name ambiguities, as a recent study (Xiao et al., 2020) pointed out that the patterns of name ambiguities in large literature databases exceed those represented in a small dataset. Due to the limited name patterns, a small dataset will restrict the exploration of some data-driven techniques. Note that, although some datasets such as GESIS-DBLP[6], SCAD-zbMATH (Müller et al., 2017), and Kim-PubMed (Kim & Owen-Smith, 2021) have decent numbers of instances, they are limited in scopes (covered domains). For example, SCAD-zbMATH is designed specifically for a mathematical domain database, zbMATH[7]. Such domain-specified datasets pose a frequently encountered problem in machine learning (ML): a model trained for a domain may be unsuitable for application to another domain because different domains cover different scopes of knowledge.

Among all the datasets, WhoisWho (Xiao et al., 2020) is the one that not only has a decent size but also covers a wide domain (large scope). However, a prominent drawback with the dataset is that it shows clear biases with respect to author ethnicity and name variation, which we refer to as the third limitation. Specifically, in WhoisWho, most last names (53 out of 65) are Chinese last names, which is inconsistent with the fact that the authors are from all over the world. Note that this issue is nontrivial because many studies have confirmed that different ethnicities have different levels of ambiguities (Louppe et al., 2016), and, based on this idea, some ethnicity-based disambiguation methods have been successfully developed (Kim et al., 2021; Louppe et al., 2016; Subramanian et al., 2021). Name variation is another frequently ignored aspect in building AND datasets. Author names presented in literature databases may differ from their actual names for many reasons (Gomide et al., 2017). This issue is known to AND researchers, but remains unsolved because addressing this issue is still very challenging; some studies have pointed out this issue in the research limitations (Zhang et al., 2021) or mentioned it in relation to future works (Sanyal et al., 2021). Unfortunately, existing datasets, including WhoisWho, can scarcely touch this issue or represent this aspect adequately. For WhoisWho, the variation degree of the last names is

0.37%[8], which we believe is lower than that of real literature databases (see the Result section for a formal investigation of the problem).

In response to these limitations, we propose a practical method to automatically build labeled datasets for author name disambiguation by leveraging the two large authoritative resources, ORCID and DOI. ORCID is dedicated to reducing the risk of errors in professional-related resources by providing a persistent identifier (ORCID iD) that authors can control and manage[9]. DOI is a persistent interoperable identifier for digital objects and is developed for use on digital networks[10]. The two academic resources have hosted a considerable number of records, which provide valuable labeled information as well as sufficient research materials to construct a large labeled AND dataset that integrates more complicated name patterns. Such a high-quality dataset has the potential to meet the requirement of building more effective disambiguation methods.

## DATASET BUILDING APPROACH

### Method overview

We developed an automatic method to build our AND datasets based on the ORCID database and a literature database, MAG[11]. The reasons for choosing MAG are as follows. First, MAG has a high demand for disambiguated authors because MAG has been used in many applications. Some well-known examples are Bing, Cortana, Microsoft Word, and Microsoft Academic[12]. Second, MAG is a heterogeneous graph containing a variety of publication-relevant metadata such as citation networks, and institutions, which may be useful for the development of disambiguation methods.

The main steps of our method are shown in Figure 2. In the database layer, we retrieved the ORCID data (baseline version of October 2020) and the MAG data (baseline version of March 2019) from the respective repositories[13][14]. Then, in the second layer, we extracted those metadata that are only related to the final AND datasets from the two databases in order to reduce the storage and computational overhead. For the literature database, we extracted DOI and other citation-related metadata required by disambiguation approaches such as title and venue. For the ORCID data, we extracted author-related metadata such as ORCID iD and the authors' credible full names (CFNs) shown on the ORCID pages, as well as the DOIs of citations claimed by the authors. In the database linking layer, we employed the DOIs to connect the two databases. As DOI is an ambiguity-free indicator for digital objects, it is, therefore, safe to connect the databases. However, the ORCID system does not specify the positions of a user (author) in the claimed citations, meaning that the author-level metadata such as affiliation that are frequently used in prior AND studies can not be obtained from the linked MAG citations if the positions are not identified. To address this issue, we designed an algorithm to identify the author positions, which corresponds to the fourth layer. In addition, in this layer, we also investigated the name variation problem. In the last layer, we conducted several aggregation operations on the ORCID-MAG linked data to build our AND datasets. The following subsections elaborate on the key steps of the method.

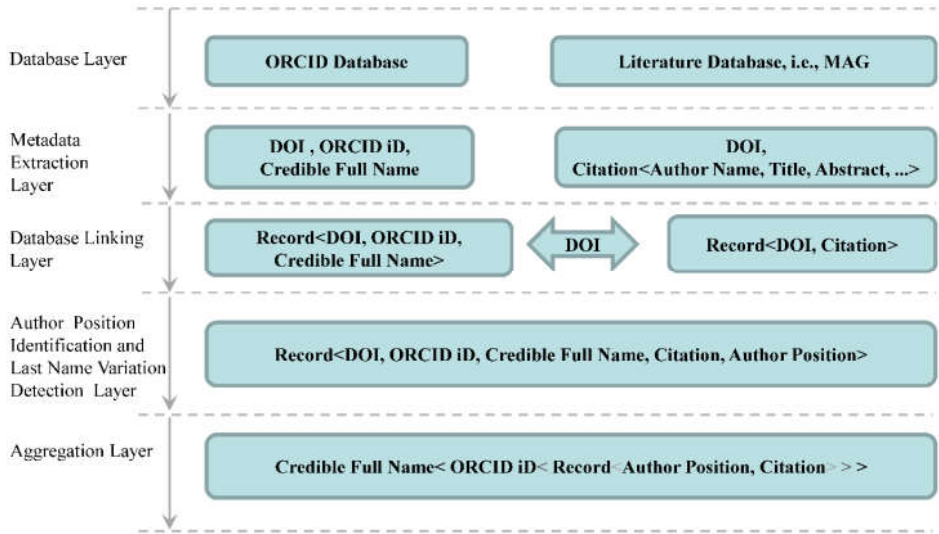

Figure 2. Dataset building pipeline

## Author position identification

The heuristic algorithm for identifying author positions is shown in Algorithm 1. For the two inputs, a CFN and the $n$ author names $[FN_1, FN_i, \ldots, FN_n]$ of a citation, the algorithm firstly maps the CFN and a $FN_i$ to the character-level 2-grams features $CFN^{(f)}$ and $FN_i^{(f)}$ respectively. For example, "John Smith" is represented by the 2-grams list of $[jo, oh, hn, \ldots, th]$. The reason for using the 2-grams measure is that it not only considers the order of characters but also is insensitive to name variants (e.g., reversed author names). Thus, the measure can be used to detect name variants. Then, the algorithm calculates the similarity $S_i$ of the CFN and $FN_i$ by measuring the number of intersections between $CFN^{(f)}$ and $FN_i^{(f)}$ to the length of the concatenated 2-grams lists. Afterward, for the citation with $n$ authors, the algorithm sorts the original author positions $I = [1, \ldots, n]$ by the corresponding similarity scores $S = [S_1, \ldots, S_n]$ in descending order, and the first element $P1 = I_1^{(r)}$ of the ranked author positions $I^{(r)}$ is likely to be the correct position. Note that, in some citations where more than one FN appears to be similar to a CFN, the algorithm may tend to incorrectly identify the author positions. For example, the two similar names "M.C. Ciornei" and "F.C. Ciornei" in the MAG citation (article ID: 2742497971)[15] have the same similarity score of 0.44 as compared to the CFN "Florina Carmen Ciornei". To handle this issue, the algorithm tries to exclude such citations by applying the rule: $P1 = I_1^{(r)}$ can be considered as the final author position only if $S_{p1}$ is higher than the second-best similarity score $S_{p2}$ ($P2 = I_2^{(r)}$, if available) by a threshold. In this study, the threshold was empirically determined to be 0.2.

**Algorithm 1**. A heuristic algorithm for author position identification

```
Input: ORCID credible full name, CFN
       Author names list of a MAG citation, FNs = (FN₁, ..., FNₙ)
Output: Identified author position, P
```
$CFN^{(f)}$ = 2-gram(CFN)
$I$ = ()
$S$ = ()
For each $FN_i$ ∈ FNs Do
    $FN_i^{(f)}$ = 2-gram($FN_i$)
    $s_i$ = 2 * intersection($CFN^{(f)}$, $FN_i^{(f)}$) / (|$CFN^{(f)}$| + |$FN_i^{(f)}$|)
    $I$ ← $i$
    $S$ ← $s_i$
End

// sorts the original author positions $I$ by the similarity *scores S* in descending order,
// and return the ranked author *positions* $I^{(r)}$.
$I^{(r)}$ = sort_index($I$, $S$)
$P1 = I_1^{(r)}$
$P2 = I_2^{(r)}$
If $S_{P1} - S_{P2} > 0.2$ Then
    $P = P1$
Else Then
    // 0 means that the author's position could not be identified.
    $P = 0$
End

return $P$

### Last name variation detection

Through author position identification, we obtained a large number of CFN-FN matches, which were used to detect name variants. Among all kinds of name variations, the last name variation is the most influential one because it is used to create the last name (LN)-based or last name and first initial (LNFI)-based blocks, which is a widely used disambiguation framework in AND studies (Levin et al., 2012; Louppe et al., 2016; Schulz, 2016) and even in production environments (Kim et al., 2016; Torvik & Smalheiser, 2009). By assuming that author names are consistent in all the authored publications, the ambiguous authors are grouped into a particular block, and they are only compared within the block. Therefore, the framework can reduce the computational complexity. However, this assumption is idealized because there are many kinds of name discrepancies resulting from various reasons (see supplemental material B). Based on the analysis, the name variation problem will eventually result in a performance reduction as the citations of the same author may be divided into different blocks.

In view of this, we herein show how to detect the last name variants and measure the degree of the variation. Specifically, we compared the last names recorded in literature databases to the authors' official last names shown on the ORCID pages. It should be noted that many literature databases such as MAG do not provide the last name field[16], making the name comparison unfeasible. We developed three measures to address this issue. The first one is "Endwith", representing whether an FN *ends with* the credible last name. The second measure extracts the last names from FNs using Joshfraser[17], which is a popular name parser working with complex, language-independent names. The criterion for determining a name variant is whether the extracted last name is identical to the ORCID last name. The third measure follows the same criterion but adopts another name parser, Derek73[18]. This tool has attracted many developers to continuously improve for it over 10 years and had been used by over 700 applications as of May 2022.

### Block-based AND dataset building

We built our block-based AND dataset (LAGOS-AND-BLOCK) with Algorithm 2. Based on the connected database $DB_{ocbib}$ between the ORCID database $DB_{oc}$ and the literature database $DB_{bib}$ (i.e., MAG in this study), we aggregated the connected citations at the author level and then at the block level to build the dataset.

**Algorithm 2**. An algorithm for automatically building LAGOS-AND-BLOCK

```
Input: ORCID database, DB_oc
       A literature database, DB_bib
Output: LAGOS-AND-BLOCK
// extract required metadata from database, rec stands for a record in a database.
DBMD_oc = {<rec.DOI, rec.ORCID, rec.CFN> | rec ∈ DB_oc}
DBMD_bib = {<rec.DOI, rec.Citation> | rec ∈ DB_bib}
// linking databases
DB_ocbib = {<rec_oc, rec_bib> | rec_oc^DOI ≡ rec_bib^DOI, rec_oc ∈ DBMD_oc, rec_bib ∈ DBMD_bib}
LAGOS-AND-BLOCK = ∅
For each <DOI, ORCID, CFN, Citation> ∈ DB_ocbib Do
      // identify the author position using Algorithm 1.
      P = AuthorPostionIdentification(CFN, Citation.AuthorNameList)
      Item = (DOI, ORCID, CFN, Citation, P)
      // find the block where this Item should be merged into
      BLK = {blk | blk.CFN ≡ CFN, blk ∈ LAGOS-AND-BLOCK}
      // find the citation group where this Item should be merged into
      CG = {cg | cg.ORCID ≡ ORCID, cg ∈ BLK}
      // update this citation group
      CG = CG ∪ {Item}
      // update this block
      BLK = BLK ∪ {CG}
      // update the dataset
      LAGOS-AND-BLOCK = LAGOS-AND-BLOCK ∪ {BLK}
End

return LAGOS-AND-BLOCK
```

At the author level, we aggregated those citations belonging to the same author into a citation group (CG) by ORCID iD. This exercise aims to restore the publication history of authors unambiguously[19][20]. At the block level, we further aggregated CGs into blocks by CFNs so that a specific block could contain multiple CGs. It is important to note that, instead of the commonly used LNFIs or FNs, we used CFNs to group the CGs because the method of building the block-based dataset has the following advantages. First, CFN is more authoritative in terms of representing blocks than LNFI or FN as the CFNs are maintained by the authors and are displayed directly on the ORCID pages without changes. In contrast, the author names presented in literature databases are error-prone. Second, it is more meaningful to disambiguate on a full-name-based dataset. In LN-based or LNFI-based datasets, authors who are apparently different persons may exist in a block. For example, the two different authors with the different names "Richard Freyman" and "Robin Freyman" can exist in the LNFI block "Freyman_R." Therefore, disambiguating on LN-based or LNFI-based datasets may yield optimistic results. In comparison, the blocks of our dataset are represented by full names, meaning that all ambiguous authors included in a particular block have the same name. This design makes our dataset more meaningful to disambiguate. Third, as shown in Figure 3, the created dataset considers synonymous names and homonymous names simultaneously. On the one hand, the author names shown in a CG may be different from a CFN, and therefore they constitute the synonymous names. On the other hand, a block usually consists of multiple CGs; thus, the names across different CGs but within the same block constitute the homonymous names.

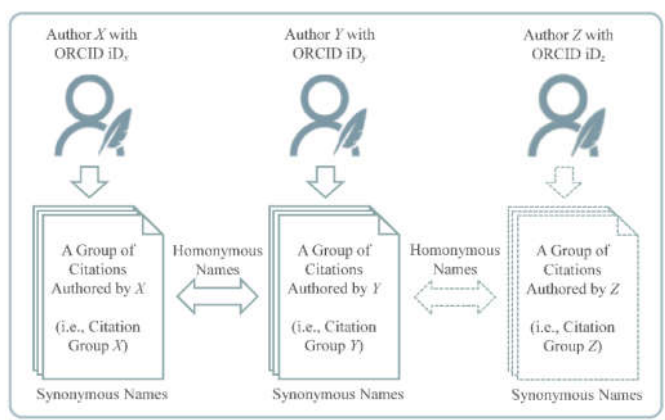

Figure 3. The structure of the block-based dataset

### Pairwise-based AND dataset building

Existing AND datasets are either block-based or pairwise-based, both of which are important because they play different roles in AND research: the block-based datasets were created for clustering-based disambiguation approaches, and the pairwise-based datasets were created for classification-based disambiguation approaches. In this study, we also considered the classification-based evaluation scenario, we created a paired-citation-based AND dataset (LAGOS-AND-PAIRWISE) based on our LAGOS-AND-BLOCK dataset following the idea of randomly sampling paired citations over blocks (Song et al., 2015; Zhang et al., 2021). Similar to LAGOS-AND-BLOCK, LAGOS-AND-PAIRWISE is also a large gold-standard dataset, which is demonstrated in the Result section.

## RESULTS

### Author position identification

We test the performance of author position identification on the three databases: PubMed, MAG, and Semantic Scholar (S2)[21]. For each database, we randomly selected 2,000 matched-name instances (FNs-CFNs) and manually examined the identified author positions. As shown in supplemental material C, the accuracies on PubMed, MAG, and S2 are 100%, 99.95%, and 99.80% respectively, demonstrating the reliability of the method.

**Last name variation**

Based on a considerable number of FN-CFN matches, we calculated the degree of variation in last names with the three mentioned measures. It should be pointed out that name variation is more pronounced for languages using some non-Western characters (e.g., "á") (Müller et al., 2017). To reflect this problem comprehensively, we reported both the character-sensitive variation degree (CSVD) and the character-insensitive variation degree (CIVD), the latter was achieved by transliterating the special characters into the standard characters (e.g., "á"→"a").

The results are presented in Table 1. We observed that the results yielded by Joshfraser and Derek73 are very similar and both are higher than the degree achieved by Endwith on MAG and S2. Although the tools may introduce parsing errors, they are indeed useful for developing AND methods because most methods rely on an explicit first name or last name field for feature computation (Wu et al., 2017) and name instance blocking (Kim, 2018; Kim et al., 2016). However, many databases such as MAG do not provide such fields. In addition, Endwith yields a CSVD of 9% and a CIVD of 6% on the three databases, such high degrees demonstrate that the last name variation problem is nontrivial in literature databases. To facilitate a better understanding, we have manually examined name variants in an attempt to summarize the typical types of variation and identify the possible reasons. From supplemental material B, we found that there can be many reasons for the name discrepancies. The provided reasons explain why last name variation is prevalent in the literature databases.

Table 1. The variation degrees of last names in three large literature databases

| Database | # Citations (C)<br># Authors (A)<br># Linked Authors (LA) | Measure | # Variants | CSVD (%) | CIVD (%) |
|---|---|---|---|---|---|
| PubMed | C: 30,128,785<br>A: 121,251,488<br>LA: 6,082,042 | - | 489147 | 8.04 | 5.80 |
| MAG | C: 213,972,535<br>A: 561,517,211<br>LA: 12,613,771 | Derek73 | 1469738 | 11.65 | 9.05 |
|  |  | Joshfraser | 1499691 | 11.91 | 9.33 |
|  |  | Endwith | 1170892 | 9.28 | 6.34 |
| S2 | C: 179,590,271<br>A: 476,379,238<br>LA: 13,697,566 | Derek73 | 1691295 | 12.35 | 9.36 |
|  |  | Joshfraser | 1718657 | 12.55 | 9.59 |
|  |  | Endwith | 1302345 | 9.51 | 6.13 |

## Multi-faceted evaluation of LAGOS-AND

We present evidence to demonstrate that the two LAGOS-AND datasets can be regarded as standard resources for author name disambiguation. For this purpose, we performed an evaluation to present the closeness between LAGOS-AND and the whole MAG in multiple facets. Note that the evaluation was performed after pruning those citations that are over-presented on a certain facet from our datasets to approximate the real distribution of MAG in that facet.

Before conducting the multi-faceted evaluation, we show the accuracy of authorship in the generated LAGOS-AND datasets. To do this, we randomly selected 1,000 instances (paired citations) from LAGOS-AND-PAIRWISE and determined the authorship of the paired citations manually. The results show that the accuracy is 99.7% (only three errors were found), demonstrating that our datasets are very accurate in terms of labeled authorship.

### Last name variation

We used the "Endwith" measure to calculate the variation degree of the last name for our datasets. The CSVD and CIVD for LAGOS-AND-BLOCK were identified at 9.63% and 6.46% respectively; and for LAGOS-AND-PAIRWISE, the CSVD and CIVD were identified at 9.72% and 6.55% respectively. Because the variation degrees are very close to the degrees of MAG (9.28% and 6.34% shown in Table 1), the two LAGOS-AND datasets are able to represent MAG in terms of this aspect.

### Publication date distribution

Figure 4a shows the number of publications each year. In comparison to MAG, LAGOS-AND reflects the tendency of the number of publications before 2010; however, the two curves of LAGOS-AND grow faster than MAG after 2010[22]. We attribute this to the creation timeframe of the ORCID system. ORCID launched its registry service in 2012[23] (see Figure 1). As a result, the papers published earlier than this timeframe may be underrepresented in ORCID. Additionally, the increasing popularity of ORCID iDs also exacerbates the under-representation problem of "older papers." A simple measure to handle this would be pruning those citations published after 2010. However, according to our experiments, such a measure would not only significantly reduce the size of our datasets but would also deprive our datasets of many valuable name ambiguity patterns, which is detrimental to the development of effective AND methods. Due to this, we decided not to make adjustments for those citations published after 2010.

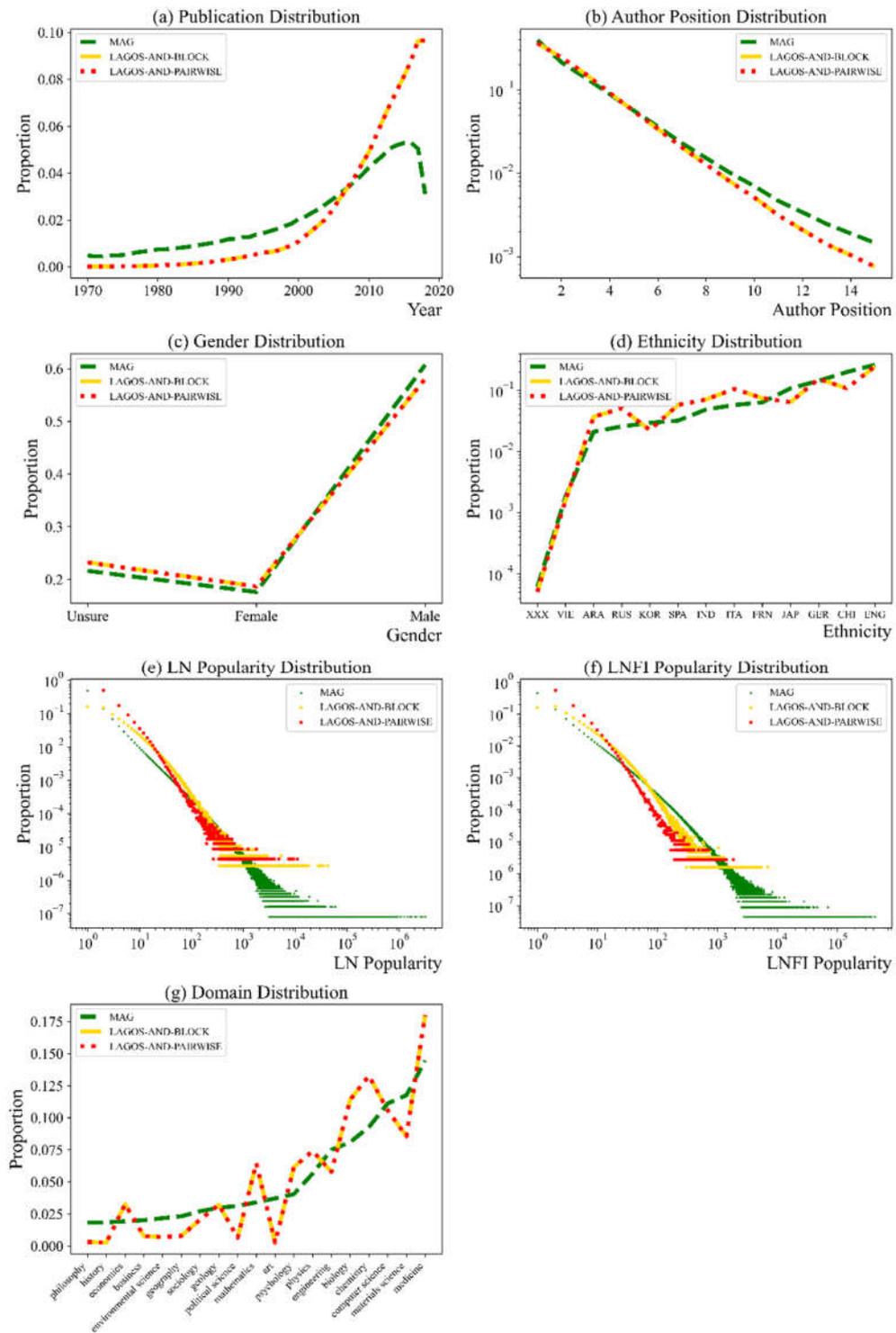

Figure 4. Multi-faceted evaluation of the LAGOS-AND datasets

### Author position distribution

Some datasets focus only on a particular author position, for example, Song-PubMed (Song et al., 2015) was created to disambiguate the first author. We note that over-focusing on a position will bias the datasets because the underlying information of the first author such as author affiliation may be richer than that of other positions in some databases (Song et al., 2015). In this study, we considered all author positions equally. As shown in Figure 4b, the closeness of the three curves demonstrates that our datasets can represent the whole MAG in this aspect.

### Gender distribution

Gender distribution is an important facet to examine the quality of AND datasets because there is a correlation between name patterns and genders (Jia & Zhao, 2019; To et al., 2020; Wais, 2016). However, this facet has not received enough attention in previous datasets. To examine the gender distribution, we used Genni+Ethnea (Smith et al., 2013; Torvik & Agarwal, 2016), a widely used gender dataset containing 4,934,974 distinct names collected from PubMed (Kim & Owen-Smith, 2021; Subramanian et al., 2021). We queried genders from Genni+Ethnea by author names to obtain the gender predictions of LAGOS-AND and MAG. As shown in Figure 4c, the closeness of the curves suggests that LAGOS-AND can represent MAG in terms of gender distribution.

### Ethnicity distribution

Person names of different ethnicities usually have different levels of ambiguities. For instance, Chinese authors are more difficult to disambiguate than other ethnicities (Gomide et al., 2017; Kim et al., 2019; Kim & Diesner, 2016). Here, we present an ethnicity distribution to demonstrate that there is no significant bias in LAGOS-AND. Similar to gender detection, detecting ethnicity from names also has a high confidence level because person names are highly culturally related, and many ethnicities have their own naming conventions (Treeratpituk & Giles, 2012). Specifically, we used an ethnicity prediction dataset EthinicSeer, a part of Genni+Ethnea, to associate the ethnicity predictions to the name instances of LAGOS-AND and MAG[24]. Because the three curves shown in Figure 4d are very close, our LAGOS-AND datasets can therefore represent the whole MAG in this aspect.

### Name popularity distribution

Another way to examine how our dataset represents MAG is to compare the name popularity, defined as the frequencies of a name in a database. We considered two kinds of name popularity, LN popularity and LNFI popularity, the proportions of which are illustrated in Figure 4e and Figure 4f respectively. From the results, we found that the minimum percentages of the two LAGOS-AND curves are different from that of MAG. The reason is that MAG covers a wider range of name popularity (LN: [1-3193636], LNFI: [1-424493]) in comparison with the name popularity of LAGOS-AND-BLOCK (LN: [1-41764], LNFI: [1-6945]) and the name popularity of LAGOS-AND-PAIRWISE (LN: [2-11096], LNFI: [2-1864]). Therefore, the minimum percentages of LN and LNFI popularity in the three databases (MAG LN: 8.03e-8, LAGOS-AND-BLOCK LN: 2.75e-6, LAGOS-AND-PAIRWISE LN: 4.28e-6; MAG LNFI: 4.45e-8, LAGOS-AND-BLOCK LNFI: 1.62e-6, LAGOS-AND-PAIRWISE LNFI: 2.74e-6) are different. Despite the dissimilarities, the tendency of the three curves is similar, which demonstrates that our datasets can overall represent the whole MAG in this facet.

### Domain distribution

Another prominent difference between our dataset and others is domain coverage. To reflect this, we adopted the level-0 field of study (FoS) of MAG, a set of 19 concepts for classifying the full disciplines of science (Shen et al., 2018), as a proxy to describe the domain distribution. From the close similarities of the curves in Figure 4g, we made the inference that LAGOS-AND not only covers a variety of domains but is also representative of the whole MAG in this facet.

## DISAMBIGUATION METHODS EVALUATION ON LAGOS-AND

In this section, we provide an evaluation of several disambiguation methods and the MAG's author ID system on the two LAGOS-AND datasets, which will serve as baselines for future AND studies interested in LAGOS-AND.

### Evaluation datasets

We used the LAGOS-AND-BLOCK dataset to evaluate clustering-based AND methods and used the LAGOS-AND-PAIRWISE dataset to evaluate classification-based AND methods. To facilitate the development of disambiguation methods, we split the two datasets into training, validation, and test folds following the same ratios of 50:25:25 (for LAGOS-AND-BLOCK, instance=block; for LAGOS-AND-PAIRWISE, instance=paired-citations), and aligned the two datasets in each data fold to ensure instances with the same CFN belonging only to a fixed fold. For example, the instance (MAG paper IDs:2093242633 and 2464285545) with the CFN "Paulo Silva" in the test set of LAGOS-AND-PAIRWISE can only be found in the test set of LAGOS-AND-BLOCK.

### Baseline methods

Depending on the disambiguation scenarios, the disambiguation methods have different implementations. In the classification-based scenario, the disambiguation methods will try to predict the authorship of paired citations, Here, we followed the commonly used supervised learning ideology with handcrafted features (Song et al., 2015; Zhang et al., 2021) to develop our methods on LAGOS-AND-PAIRWISE. In contrast, in the clustering-based scenario, the disambiguation methods will try to attribute citations to the right authors. Therefore, we followed the semi-supervised ideology to develop our methods on LAGOS-AND-BLOCK. The semi-supervised methods disambiguate authors by applying the supervised AND methods developed in the classification-based scenario to derive the author similarities and then using a clustering algorithm (Torvik & Smalheiser, 2009; Cota et al., 2010; Smith et al., 2013; Cen et al., 2013; Ferreira et al., 2014; Qian et al., 2015; Louppe et al., 2016) such as Hierarchical Agglomerative Clustering (HAC) to group citations into disjoint clusters based on the calculated author similarities.

The baseline methods developed in this study are simple as well as generic because the underlying metadata are available in most literature databases (see the supplemental material A), and the logic of feature extractions is straightforward. Table 2 itemizes all the features used in the baseline methods. We divide the features into two groups: the base feature group $BF$ and the content feature group $CF$ according to the metadata being used. For two particular citations, the $BF$ feature group contains four basic features: the similarity of author names, the gap years between the two publication dates, the similarity of publication venues, and the similarity of author affiliations, which are calculated by the measures shown in Table 2. The $CF$ feature group contains four kinds of content-based features extracted by different measures on the same article content (i.e., title and abstract). The reasons why we paid special attention to the content similarity are that content similarity is helpful for improving disambiguation (Kim et al., 2019) and, more importantly, authors can be intuitively disambiguated by judging the closeness of the research topic of two citations when other metadata such as affiliation are missing or the author names provide little discriminative information (e.g., the ambiguous authors share the same full name in our datasets).

As shown in Table 2, the similarity measures are Jaccard Index, TFIDF, Doc2vec (Le & Mikolov, 2014), and a simple neural network that can capture the content similarity at the semantic level (see supplemental material D for the diagram and the parameter settings of the network)[25]. Based on all the $BF$ features and a $CF$ feature, we developed several baseline methods, which can be found in Table 3 and Table 4. We evaluated these methods on both LAGOS-AND-BLOCK and LAGOS-AND-PAIRWISE datasets. Additionally, we evaluated the MAG's author ID system (denoted by MAG-Author-ID), which was created by the Microsoft Academic team for the over 560 million authorship in MAG. We evaluated MAG-Author-ID because the ID system has been widely used for many downstream tasks (Färber, 2019; Huang et al., 2020). However, it is unclear whether the actual performance of the ID system represents the uniqueness of the authors.

Table 2. Features list and feature extraction measures

| Feature Group | Feature Name | Metadata | Measure |
|---|---|---|---|
| Base Feature (BF) Group | Name Similarity | Full Author Name | Char-Level Jaccard Index (2-gram) |
| | Publication Year Gap | Publication Year | Absolute Difference |
| | Venue Similarity | Venue | Word-level Jaccard Index |
| | Affiliation Similarity | Affiliation | Word-level Jaccard Index |
| Content Features (CF) Group | $CF_{jaccard}$ | Title and Abstract | Word-level Jaccard Index |
| | $CF_{tfidf}$ | Title and Abstract | TFIDF |
| | $CF_{doc2vec}$ | Title and Abstract | Doc2vec |
| | $CF_{nn}$ | Title and Abstract | Neural Network |

### Metrics and parameter settings

We report precision (P), recall (R), F1, and Macro-F1 metrics for the classification-based AND approaches. The reason for using the additional metric Macro-F1 is that the LAGOS-AND-PAIRWISE is naturally skewed (95.56% of instances are positives), and Macro-F1 is a more suitable metric for evaluation on such an imbalanced dataset. It is not surprising that most instances in LAGOS-AND-PAIRWISE are positives because most blocks of LAGOS-AND-BLOCK contain citations belonging to a single author (see the supplemental material E), therefore, sampling over LAGOS-AND-BLOCK likely yields positive samples. To measure the performance of clustering-based AND approaches, we use B-cubed (B3) precision (B3-P), B-cubed recall (B3-R), and B-cubed F1 (B3-F1) as the metrics have been widely used in prior clustering-based AND studies (Han et al., 2017; Qian et al., 2015).

In terms of model settings, we used the Random Forest (RF) algorithm to predict the similarity of paired authors and used the HAC algorithm to cluster the ambiguous authors, because RF has achieved robust performance in prior studies (Sanyal et al., 2021), and HAC is also a commonly adopted clustering algorithm (Han et al., 2015; Wu & Ding, 2013). Here, the number of tree components of the RF classifiers is set to 100. Note that, in contrast to some supervised algorithms such as RF showing a robust performance, the performance of clustering algorithms is often largely affected by the built-in parameters. Thus, a tuning process should be conducted to identify the optimal clustering parameter instead of using an empirical value. For HAC, the distance threshold is the only parameter that needs to be tuned. We tuned the parameter for all the baseline methods requiring clustering on the validation set of LAGOS-AND-BLOCK by searching the parameter in the range of [0,1] with the incremental step being set to 0.05, and we determined the optimal parameter when the B3-F1 metric reached the maximal. Finally, the optimal distance thresholds of all the semi-supervised baselines were surprisingly identified at the same value 1.0, and the B3 metrics achieved by these baseline methods are the same (see Table 4). This finding implies that the B3-F1 metrics are only maximized when all the citations are merged into the same cluster. A deeper analysis suggests that these are normal behaviors, because most LAGOS-AND-BLOCK blocks consist of citations belonging to a single author (see the supplemental material E), and therefore simply merging them into one cluster would yield the best performance.

The investigation also identifies a limitation of the LAGOS-AND-BLOCK dataset, namely, it is not suitable to focus the dataset on developing clustering-based AND methods because the overwhelming "single author blocks" do not support parameter tuning for clustering algorithms. However, this does not mean that LAGOS-AND-BLOCK is completely useless. We will discuss this in the Discussion section.

To obtain meaningful clustering results, we trimmed those blocks containing only one author from LAGOS-AND-BLOCK, leaving all the blocks containing at least two real-life authors. After this step, the trimmed LAGOS-AND-BLOCK dataset (denoted by LAGOS-AND-BLOCK-TRIMMED) contains 39,528 blocks (9950 test blocks) and 758,584 citations. Although the step significantly reduced the size of LAGOS-AND-BLOCK, LAGOS-AND-BLOCK-TRIMMED is still a very large dataset: it outperforms 11 out of 12 datasets in terms of dataset size, as shown in the supplemental material A. With the trimmed dataset, we conducted the tuning process and developed our clustering-based methods on it. Finally, the optimal parameters of the baseline methods shown in Table 3 are identified at 0.45, 0.25, 0.2, 0.2, 0.25, and 0.2, respectively.

# Results

Table 3 and Table 4 show the respective evaluation results on our datasets. Our observations are as follows.

Table 3. Evaluation results on LAGOS-AND-PAIRWISE

| Method | P | R | F1 | Macro-F1 |
|---|---|---|---|---|
| Random | 95.46 | 50.01 | 65.64 | 36.9 |
| MAG-Author-ID | **98.82** | 70.16 | 82.06 | 51.12 |
| Name Similarity | 95.8 | 87.57 | 91.5 | 50.08 |
| BF | 95.55 | **99.56** | 97.51 | 50.16 |
| BF + $CF_{jaccard}$ | 95.62 | 99.31 | 97.43 | 51.53 |
| BF + $CF_{tfidf}$ | 95.67 | 98.53 | 97.08 | 52.35 |
| BF + $CF_{doc2vec}$ | 95.67 | 98.65 | 97.14 | 52.46 |
| BF + $CF_{nn}$ | 96.57 | 98.57 | **97.56** | **65.21** |

Table 4. Evaluation results on LAGOS-AND-BLOCK and LAGOS-AND-BLOCK-TRIMMED[26]

| | Method | B3-P | B3-R | B3-F1 |
|---|---|---|---|---|
| LAGOS-AND-BLOCK | MAG-Author-ID | 99.88 | 64.73 | 70.59 |
| | All learnable baselines | 97.79 | 100 | 98.52 |
| LAGOS-AND-BLOCK-TRIMMED | MAG-Author-ID | **97.68** | 71.11 | 77 |
| | Name Similarity | 70.37 | 87.63 | 74.78 |
| | BF | 75.85 | 86.62 | 77.4 |
| | BF + $CF_{jaccard}$ | 77.6 | 89.07 | 79.61 |
| | BF + $CF_{tfidf}$ | 77.27 | 90.09 | 79.93 |
| | BF + $CF_{doc2vec}$ | 74.14 | **91.62** | 78.69 |
| | BF + $CF_{nn}$ | 79.68 | 89.59 | **81.16** |

First, we found that F1 and Macro-F1 of MAG-Author-ID are 82.06% and 51.12% on LAGOS-AND-PAIRWISE, and similarly, the achieved B3-F1 score on LAGOS-AND-BLOCK is 70.59%. It is surprising to see that the performance of the disambiguated ID system is much lower than expected, given that it has been widely used by many studies (Huang et al., 2020). In addition, we found that MAG-Author-ID is high in precision but low in recall. This can be explained by the method of building the author ID system (Wang et al., 2020). The Microsoft Academic engine harvests scientific articles online, thus, it can find many personal websites and public curricula vitae containing the author's publication list. Since the author-article relationship in the publication list is very accurate, the author ID system created by the method achieved a high level of precision. However, a critical issue with the method is that the crowdsourced publication list of authors is often not complete. To deal with this issue, the research team of MAG developed a machine learning approach to merge other possible articles to the authors when the predictions by the approach exceeded a 97% confidence threshold. The method indeed improved the incompleteness, however, this conservative method inevitably split the articles by the same author into multiple clusters because a high confidence score needed to be met. This approach makes MAG-Author-ID achieve a low call.

Second, we found that combining a content-based feature with all $BF$ features significantly improved the disambiguation performance and that different content features have different contributions. The method BF + $CF_{nn}$ was proven to be the best performer, which improved BF by a wide margin. This evidence suggests that the content information is very helpful for disambiguation methods in our datasets.

## DISCUSSION

### Insights into LAGOS-AND

We performed a feature analysis to help understand the characteristics of LAGOS-AND, as feature contributions can reflect the importance of the metadata in our dataset. The feature contributions of the best-performing method BF + $CF_{nn}$ are shown in Figure 5, where the vertical lines and dots denote the standard deviations and means of the feature contributions across all RF ensemble trees. We found that the neural network-based content similarity has the highest contribution, which demonstrates that article content can be effective for disambiguation. Moreover, an interesting finding is that, in contrast to other AND datasets, the name metadata in our dataset is less discriminative. This can be understood by the fact that, in a given block, the ambiguous authors have the same full name, i.e., CFNs, and therefore very limited discriminative information can be obtained from names.

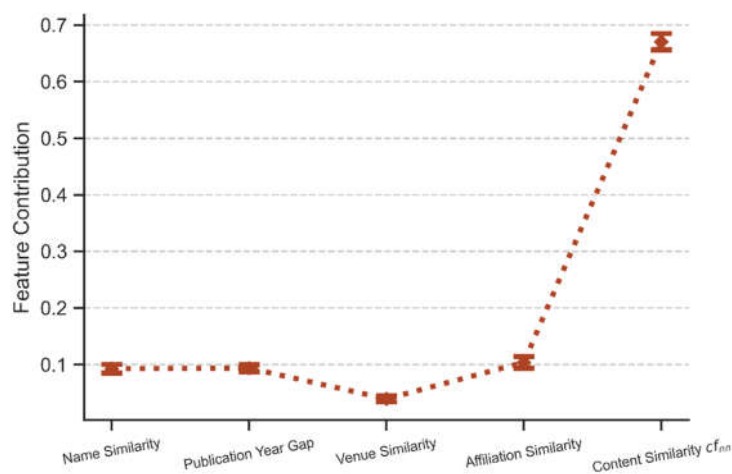

Figure 5. Feature contribution analysis for $BF + CF_{nn}$; scores are voted by the ensemble trees of the Random Forest classifier

### Drawback of LAGOS-AND

The parameter tuning process has demonstrated that the LAGOS-AND-BLOCK dataset is not suitable for developing AND methods requiring clustering. However, this does not mean that the dataset is completely useless. At least, LAGOS-AND-BLOCK provides a platform for the evaluation of disambiguated author IDs in a gold standard manner as disambiguated author IDs do not require clustering and parameter tuning. This point is important for two reasons. First, *evaluating disambiguated ID systems is as important as evaluating disambiguation methods*. Many existing methods have achieved a high-performance score of more than 90% on test datasets (Vishnyakova, et al., 2016; Zeng & Acuna, 2020). However, we argue that these methods may encounter performance reduction if they are applied to real literature databases because the blocks of whole databases are usually larger than those of the evaluation datasets. For example, the largest block, "David Smith," in our dataset contains 1,067 citations while the corresponding block in MAG contains 4,033 citations, suggesting that name disambiguation on literature databases such as MAG is much more difficult than disambiguation on the evaluation datasets. Thus, we can infer that the performance achieved by a disambiguation method is likely higher than the performance of the disambiguated author IDs created by the disambiguation method on the same test set. In this sense, an independent evaluation of disambiguated ID systems is greatly important for AND research. Second, *it is also important to use a gold standard dataset to evaluate disambiguated author IDs*. Given that existing datasets are more or less biased, the gold standard LAGOS-AND-BLOCK dataset can reflect the performance of the disambiguated author IDs in a realistic scenario.

Despite the above, we have shown that the drawback is relatively easy to overcome. By simply removing the blocks containing a single author from LAGOS-AND-BLOCK, we created another block-based dataset, LAGOS-AND-BLOCK-TRIMMED, which can be used to develop clustering-based methods. This implies that the two block-based datasets play different roles in AND studies. Specifically, we suggest that future studies that are interested in LAGOS-AND use LAGOS-AND-BLOCK to *test* the disambiguated author IDs and use LAGOS-AND-BLOCK-TRIMMED to *develop and evaluate* the disambiguation methods requiring clustering.

### Error analysis for LAGOS-AND

Although we followed rigorous procedures to build the LAGOS-AND datasets, we realize that our datasets are not error-free. Here, we summarize several reasons that could result in the errors according to our intensive observations. (1) Reversed Names. In the name management interface of the ORCID system, the input boxes of the first name and the last name are explicitly distinguished to ensure that authors (users) will enter the right name components into the boxes. Though this kind of error is extremely rare, we still observed outliers. For example, when writing this paper, we found that the author named "Ruixue, Sun" with ORCID iD "0000-0003-2495-0433" has reversed her/his name to "Sun, Ruixue" on the author's ORCID page. (2) Author with Multiple ORCID iDs. As the ORCID team claims, there might be some authors who have created multiple ORCID iDs[27]. Fortunately, the ORCID team has developed several measures to prevent such errors from occurring or to eliminate them if they do occur. For instance, when a new registration is received, the ORCID system will attempt to block the registration by searching the registration database for a matching existing account/accounts. If possible accounts are found, the system will return the alternatives to the author for selection. The system also allows authors to manage the already created duplicates in case they have been unintentionally created, i.e., marking one iD as the primary and deprecating others [28]. These measures are indeed helpful for eliminating these kinds of errors, however, we suspect that there are still undetected duplicates inside the ORCID data. (3) Incorrect Author Position Identification. The author position identification algorithm does not necessarily guarantee perfect performance. As shown in supplemental material C, the algorithm fails for 0.05% of MAG author names, and thus this step will introduce errors into our datasets.

### Implications of the performance of MAG author IDs

Our evaluation shows that the MAG's author ID system only achieved a 70.59% B3-F1 score, an 82.06% F1 score, and a 51.12% Macro-F1 score on our gold standard datasets. Such low performance may lead to distorted results for those studies drawn on this basis. In view of this, we suggest that future studies or applications should be more careful in using MAG author IDs. Additionally, we found that there is a significant gap between the performance of MAG author IDs and many disambiguation methods. For example, many methods have achieved a performance of more than 90% (Song et al., 2015; Vishnyakova et al., 2019). The discrepancies in performance highlight an important research question about the practicality of AND methods: many studies approaching AND used fancy techniques such as heterogeneous graphs (King et al., 2014) and adversarial learning (Peng et al., 2019), however, most of them are limited in terms of being used on large literature databases such as MAG and OpenAlex[29] due to the high computational complexity (Xiao et al., 2020). The lessons learned from the significant gaps will help better understand the name ambiguity problem and the performance of disambiguation methods in real-world large-scale literature databases.

### Implications of last name variation

By connecting the ORCID data to three large literature databases, the variation degrees in last names were identified at 8.04%-12.55% (CSVD) and 5.80%-9.59% (CIVD). Notably, the problem is nontrivial because it plays an important role in the widely accepted block-based disambiguation framework, in which the author's last name is assumed to be consistent across all the author's publications, and the last name (or the last name and first initial) is used to group name instances into blocks so that disambiguation for large literature databases will be more computationally efficient. However, the high variation degrees suggest that an author's publications may be divided into multiple blocks and thus assigned directly to different authors. Based on the analysis, this finding is important in revealing the limitation of the classic block-based disambiguation framework, as well as helping future studies develop a better disambiguation framework.

### Research limitations

The first limitation is the potential errors of the ORCID names (CFNs). Although the names are maintained by the authors themselves, we indeed find reversed names (very rare). However, it is difficult to detect the reversed names because determining whether author names are reversed is often confusing without strong background knowledge about the naming conventions of different groups of people. The second limitation is that we have not fully considered the ORCID iD duplicates. Although the measures provided by the ORCID team are effective to eliminate the duplicates, we failed to find a way to identify and remove all the possible duplicates. Third, we calculated the last name variation degree for three literature databases by comparing the author name instances inside these databases to CFNs. It should be pointed out that the degrees can be influenced by many factors, for example, whether authors have uploaded all their publications to the ORCID system. Unfortunately, we are unable to examine the impact of these factors because the underlying information is not available.

## CONCLUSIONS

In this paper, we described a method that can automatically build large labeled datasets for the author name disambiguation research. Based on the method and the academic resources ORCID and DOI, we built two AND datasets: LAGOS-AND-BLOCK and LAGOS-AND-PAIRWISE, which not only have a large size but also show close similarities to the whole Microsoft Academic Graph across validations of six facets. In building the dataset, we investigated the last name variation problem and revealed the variation degrees in three considerable literature databases. Furthermore, we evaluated the MAG's author ID system and several baseline methods on the created datasets; the analyses for the datasets and the experimental results are also presented in the paper.

## Supplemental Material A: A list of datasets for author name disambiguation

Table 5. A list of datasets for author name disambiguation research

| Dataset | Release Year | Structure | Containing Last Name Variants (Y/N) | Database/Domain | # Block (B) # Author (A) # Citation (C) | Available Metadata |
|---|---|---|---|---|---|---|
| Han-DBLP | 2005 | LNFI Block | N | DBLP/Computer Science | B: 14 A: 479 C: 8453 | Author Name, Coauthor Names, Paper Title, Venue |
| Qian-DBLP | 2015 | FN Block | N | DBLP/Several Domains | B: 680 A: 1201 C: 6783 | Author Name, Coauthor Names, Paper Title, Abstract, Keywords, Venue, Publish Year |
| GESIS-DBLP | 2016 | FN Block | N | DBLP/Computer Science | B: 1932 A: 5407 C: 62635 | Author Name, Coauthor Names |
| Kim-DBLP | 2018 | LNFI Block | N | DBLP/Computer Science | B: 14 A: 480 C: 4156 | Author Name, Coauthor Names, Paper Title, Venue, Publish Year |
| Tang-AMiner | 2011 | FN Block | N | DBLP/Several Domains | v1 B: 110 A: 1762 C: 8385 v2 B: 109 A: 1743 C: 8316 | Author Name, Coauthor Names, Affiliation, Paper Title, Venue, Publish Year |
| WhoisWho | 2019, 2020 | FN Block | Y (very limited number of cases) | AMiner/General Domain | v1 B:421 A: 45,187 C: 399,255 v2 B:231 A: 13,662 C: 221,802 | Author Name, Coauthor Names, Affiliation, Paper Title, Abstract, Keywords, Venue, Publish Year |
| Culotta-REXA | 2007 | LNFI Block | N | Rexa/Computer Science | B: 13 A: 329 C: 3033 | Author Name, Coauthor Names, Paper Title, Venue |
| Cota-BDBComp | 2010 | LNFI Block | N | BDBComp/Computer Science | B: 10 A: 205 C: 361 | Author Name, Coauthor Names, Venue |
| SCAD-zbMATH | 2017 | LNFI Block | N | zbMATH/MATH | B: 2919 A: 2946 C: 33810 | Author Name, Paper Title, Venue, Publish Year |
| Song-PubMed | 2015 | LNFI Block | N | PubMed/Biomedicine | B: 36 A: 385 C: 2,875 | Author Name, Author Position, PMID (for associating metadata from PubMed) |

| Dataset | Release Year | Structure | Containing Last Name Variants (Y/N) | Database/Domain | # Block (B) # Author (A) # Citation (C) | Available Metadata |
| --- | --- | --- | --- | --- | --- | --- |
| GS-PubMed | 2019 | Pairwise | Y (only 63 out of 3802 cases) | PubMed/ Biomedicine | C: 3756 | Author Name, PMID (for associating metadata from PubMed) |
| Kim-PubMed | 2021 | LNFI Block, Pairwise | Y (containing only non-ASCII-based name variants) | PubMed/ Biomedicine | AUT_ORC B: 197379 A: 245754 C: 3076501 AUT_NIH B: 29243 A: 34206 C: 312951 AUT_SCT C: 1680310 | Author Name, PMID for associating metadata from PubMed, Author Position, ORCID iD/NIH ID, ORCID Name/NIH Name, Authority2009 ID (disambiguated IDs for PubMed 2009), Publish Year, Predicted Ethnicity, Predicted Gender |

## Supplemental Material B: The types of last name variation

Many factors can result in the name discrepancies. To clarify, we compared the author names of MAG with the corresponding credible names shown in the ORCID system. Several types of last name variations with possible reasons are summarized in Table 6.

Table 6. The types of last name variations. The example names on the left side of the parenthesis are from MAG and the names inside the parenthesis are from ORCID

| Variation Type | Examples | Possible Reasons |
| --- | --- | --- |
| Reversed Name | Wang Fan (Fan, Wang) Chang Rui (Rui, Chang) Acosta Manuel (Manuel, Acosta) | (1) Different name writing conventions in different cultures and ethnic groups. |
| Totally Changed Last Name | Monika Wnuk (Monika, Friedemann) Preety Ramful (Preety, Srivastava) Rocio M. Uresti (Rocio, Margarita) | (1) Last name changed when female authors get married. (2) Incorrect name recognition during parsing from the published paper. |
| Misspelling of Last Name | Remko Leys (Remko, Leijs) Sean o Duill (Sean, o'duill) Ansgar Höper (Ansgar, Hoeper) | (1) Missspelling by the authors. (2) Incorrect name recognition during parsing from the published paper. (3) Author's preference: an author may use different names to trace different publications. |
| Sticky Last Name. Consecutive author names incorrectly stick together | Ana Charas and Jorge Morgado, (Ana, Charas) Basudeb Basu and Bablee Mandal (Bablee, Mandal) Mert M. Oymak and Deniz Uner (Mert Mehmet, Oymak) | (1) Incorrect name recognition during parsing from the published paper. |
| Incomplete Last Name. Usually consists of more than one name components | M.A. Mohammed (Mahmoud, Mohammed Mahros) Paule Martel (Paule, Latino-martel) L.S. Dias (Luís, Silva Dias) | (1) Incorrect name recognition during parsing from the published paper. (2) A part of the name components is ignored by the authors. |

## Supplemental Material C: An evaluation of author position identification

To evaluate the author position identification algorithm, we applied the algorithm to three large literature databases: PubMed, Microsoft Academic Graph, and Semantic Scholar. We randomly selected 2,000 instances from the matched name instances and manually examined whether the positions were correctly identified. As shown in Table 7 below, the high accuracies suggest that this algorithm could effectively identify the author positions for a given ORCID name from citations.

Table 7. Evaluation of the author position identification algorithm on three large literature databases based on 2,000 randomly sampled instances

| Databases | # Errors | Accuracy (%) |
|---|---|---|
| PubMed | 0 | 100.00 |
| MAG | 1 | 99.95 |
| S2 | 4 | 99.80 |

**Supplemental Material D: A simple neural network for capturing content similarity of citations**

Figure 6 shows the neural network model used to capture the content similarity of citations at a semantic level. The model accepts the content information (i.e., title and abstract) of paired citations, then the content words are mapped to embeddings via a pre-trained word embedding model GloVe (Pennington et al., 2014). A Bi-directional Gated Recurrent Unit (BiGRU) (Cho et al., 2014) layer is used to learn hidden information from the input sequences. Finally, the outputs of the two-sided BiGRU layers are concatenated and fed to the Multiple Layer Perceptron (MLP) layers, which map the concatenated vector to the content similarity. For parameter settings, we truncated the maximum length of input to 300. The number of MLP layers was set to 3, and the learning rate was set to 5e-5. We used the Adam optimizer to automatically adjust the learning rate during training. The training epoch was set to 30 while saving the best model with the lowest loss on the validation set.

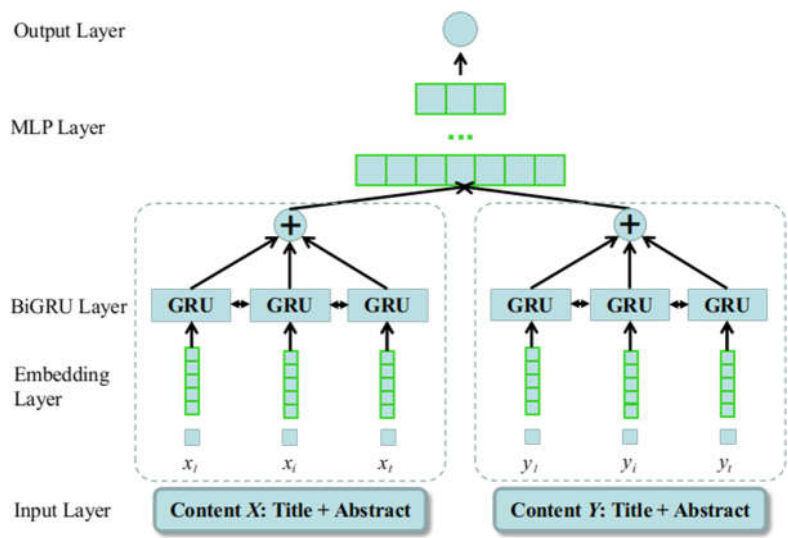

Figure 6. A simple neural network for measuring the content similarity of two citations

## Supplemental Material E: Visualization of LAGOS-AND-BLOCK characteristics

Because the structure of LAGOS-AND-BLOCK is more complex than that of LAGOS-AND-PAIRWISE, we show the characteristics of LAGOS-AND-BLOCK with respect to block size distribution and the distribution of last name variants.

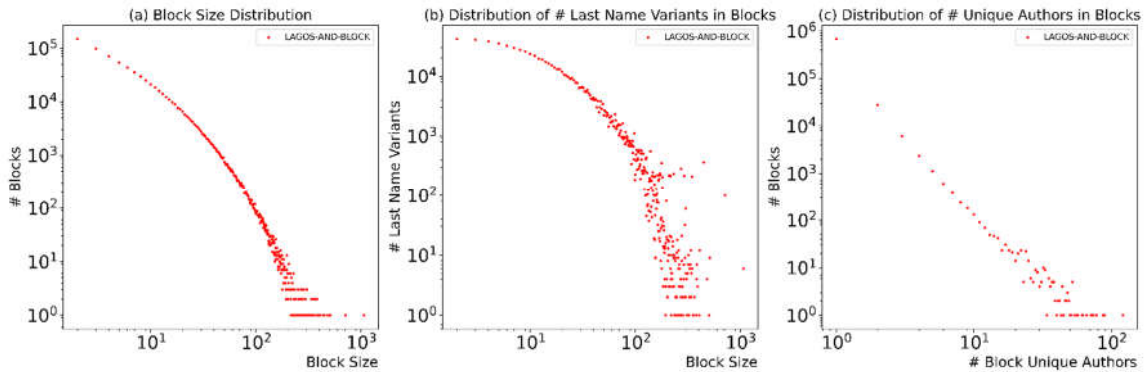

Figure 7. Visualization of LAGOS-AND-BLOCK characteristics

The block size distribution represents the number of citations within a block. As shown in Figure 7a, the gold standard reveals the fact that the real-word author name ambiguities are distributed in a wide range of block sizes [2-1067], and that most of the ambiguous names appear in small blocks, which is different from some datasets, such as WhoisWho (Xiao et al., 2020) that focus on large blocks only (i.e., popular author names). Moreover, we explore the relationship between the number of last name variants and block size to demonstrate how our dataset represents the problem of last name variation. As shown in Figure 7b, we found that most variants occur in small blocks, which seems contrary to the intuition of larger blocks having more name variants. This finding can be explained by the fact that larger blocks often refer to popular names, which tend to be shorter, and therefore they are less likely to be varied. For example, in our block-based dataset, the average length of the last names for the 100 biggest blocks is 3.73, while it is 7.28 for the 100 smallest blocks. Since different blocks cover different numbers of real authors, Figure 7c reveals the distribution of the number of authors in a block, with the horizontal axis being the number of unique authors in a block and the vertical axis being the number of corresponding blocks in LAGOS-AND-BLOCK. We observed that there are 685,547 blocks containing only one author, which constitutes 94.5% of all blocks in LAGOS-AND-BLOCK.

## ACKNOWLEDGEMENTS

This work was supported by the National Key Research and Development Program of China [No.2019YFB1404702]. The authors would like to express their sincere gratitude to the editors and the anonymous reviewers for their valuable comments and suggestions.

## AUTHOR CONTRIBUTIONS STATEMENT

**Li Zhang**: Conceptualization, Investigation, Methodology, Software, Formal analysis, Resources, Data curation, Writing - Original Draft, Review & Editing. **Wei Lu**: Supervision, Project administration, Funding acquisition, Methodology, Review & Editing. **Jinqing Yang**: Language Editing.

## DATA AVAILABILITY STATEMENT

The initial versions of our datasets are available at https://zenodo.org/record/4568624, and the code for the baseline methods as well as the analysis presented in this paper is available at https://github.com/carmanzhang/LAGOS-AND.

## ORCID

Li Zhang: https://orcid.org/0000-0003-2104-0194

## CONFLICT OF INTEREST

None declared.

## ENDNOTES

[1] In this study, we use *citation* rather than *paper* or *article* to represent the published papers, as most literature databases only contain article metadata.
[2] https://www.aminer.cn/
[3] https://orcid.org
[4] https://www.doi.org
[5] Note that we eliminated the KISTI-AD-E-01 dataset created by (Kang et al., 2011) from this review because it is not retrievable according to the given link.
[6] https://doi.org/10.7802/1234
[7] https://www.zbmath.org
[8] Note that WhoisWho contains two name writing styles because a large number of citations are Chinese citations in which Chinese authors prefer to write their last name first. To accurately calculate the degree of variation for WhoisWho, we identified the

Chinese papers and converted the writing style of the Chinese names to the standard Western name style.

[9] https://info.orcid.org/researchers/
[10] https://www.doi.org/
[11] Note that all experimental results reported in this paper are based on the version v1.0 of the LAGOS-AND datasets. By rerunning our dataset creation pipeline on the OpenAlex database, we have created the second version of the datasets, available at https://zenodo.org/record/6731767#.YrhAJHVBw5k. We built them on OpenAlex instead of MAG because MAG was discontinued on 31 December 2021, and OpenAlex not only positions itself as a drop-in replacement for MAG but also keeps evolving by aggregating academic resources from other repositories. We plan to release the third versions in 2023-2024.
[12] https://www.microsoft.com/en-us/research/project/microsoft-academic-graph
[13] https://orcid.figshare.com/articles/dataset/ORCID_Public_Data_File_2020/13066970/1
[14] https://zenodo.org/record/2628216#.YBI2KtUzaUk
[15] Only first name initials are available in this citation.
[16] There is no explicit field for last name in MAG, only full names are available.
[17] https://github.com/joshfraser/PHP-Name-Parser
[18] https://github.com/derek73/python-nameparser
[19] https://members.orcid.org/api/tutorial/reading-xml
[20] https://support.orcid.org/hc/en-us/articles/360006973853
[21] https://www.semanticscholar.org/
[22] Note that the MAG curve declines after 2018, which probably caused by the incomplete indexing of citations published after 2018.
[23] https://info.orcid.org/orcid-launches-registry
[24] The ethnicity predictions included in EthnicSeer are Vietnamese (VIE), Arabian (ARA), Russian (RUS), Korean (KOR), Columbian-Spanish-Venezuelan (SPA), Indian (IND), Italian (ITA), French (FRN), Japanese (JAP), German (GER), Chinese (CHI), British (ENG), and others (XXX).
[25] It should be noted that we also tried to incorporate the handcrafted features into the neural network. However, we did not obtain a better result.
[26] Note that the "Random" baseline is not applicable to the clustering-based evaluation due to the method can not assign ambiguous authors to a specific cluster before knowing the number of the unique authors (clusters) in a block.
[27] https://info.orcid.org/managing-duplicate-orcid-ids/
[28] https://support.orcid.org/hc/en-us/articles/360006971593-Do-you-have-more-than-one-account-
[29] https://openalex.org/